\documentclass[runningheads]{llncs}
\usepackage{multirow}
\usepackage{graphicx}

\begin{document}
\title{Shared Boundary Interfaces: can one fit all? A controlled study on virtual reality vs touch-screen interfaces on persons with Neurodevelopmental Disorders}
\titlerunning{Shared Boundary Interfaces: can one fit all?}
%
\author{ Francesco Vona \inst{1}\orcidID{0000-0003-4558-4989} \and Eleonora Beccaluva \inst{1}\orcidID{0000-0001-6566-0905} \and
Marco Mores\inst{2}\orcidID{0000-0002-3046-1145} \and Franca Garzotto \inst{1}\orcidID{0000-0003-4905-7166}}
\authorrunning{F. Vona et al.}
%
\institute{
Department of Electronics, Information and Bioengineering, Politecnico di Milano, Milan, Italy \email{name.surname@polimi.it} \url{https://i3lab.polimi.it/} 
\and
Fraternità e Amicizia Soc. Coop. Soc. Onlus, Milan, Italy  \email{name.surname@fraternitaeamicizia.it} \url{https://www.fraternitaeamicizia.it/}
}
\maketitle              

\begin{abstract}
Technology presents a significant educational opportunity, particularly in enhancing emotional engagement and expanding learning and educational prospects for individuals with Neurodevelopmental Disorders (NDD). Virtual reality emerges as a promising tool for addressing such disorders, complemented by numerous touchscreen applications that have shown efficacy in fostering education and learning abilities. VR and touchscreen technologies represent diverse interface modalities. This study primarily investigates which interface, VR or touchscreen, more effectively facilitates food education for individuals with NDD. We compared learning outcomes via pre- and post-exposure questionnaires. To this end, we developed GEA, a dual-interface, user-friendly web application for Food Education, adaptable for either immersive use in a head-mounted display (HMD) or non-immersive use on a tablet. A controlled study was conducted to determine which interface better promotes learning. Over three sessions, the experimental group engaged with all GEA games in VR (condition A), while the control group interacted with the same games on a tablet (condition B). Results indicated a significant increase in post-questionnaire scores across subjects, averaging a 46\% improvement. This enhancement was notably consistent between groups, with VR and Tablet groups showing 42\% and 41\% improvements, respectively.

\end{abstract}
\section{Introduction}
Neurodevelopmental disorders (NDD) are a group of conditions that typically manifest during childhood\cite{american2013diagnostic}. Common diagnoses within this category include ASD -Autism Spectrum Disorder, ADHD -Attention Deficit and Hyperactivity Disorder, and ID - Intellectual Disability \cite{american2013diagnostic}. Children with these conditions often require support in developing cognitive skills such as attention and language, social abilities like interacting with others, and personal and household independence skills. Concerning the prevalence of these disorders, precise data is elusive, yet there is a consensus on its increasing occurrence. The World Health Organization (WHO) reports that ASD affects one in every 160 children \cite{wing1993definition}. 
In numerous neurodevelopmental disorders, the significance of food management and, consequently, food education cannot be overstated, acting as a pivotal component in the overall treatment and quality of life improvement for affected individuals \cite{MATSON2001165}. The prevalence of obesity among individuals with Neurodevelopmental Disorders (NDD) is notably higher compared to the general population. This increased prevalence is attributed to a combination of factors including genetic predispositions, medication side effects, reduced physical activity, and dietary challenges. 
In this context, diverse eating disorders have been noted. Sisson and Van Hasselt \cite{sisson1989feeding} detailed four general types of feed behaviors typical of NDD:
\begin{itemize}
    \item Challenges in eating independently
    \item Disruptive eating behaviors
    \item Inconsistent eating volume (over or under-eating)
    \item Food pickiness
\end{itemize}

These classifications, however, do not cover all scenarios. Matson and Bamburg \cite{matson1999descriptive} discovered a significant prevalence (9-25\%) of Pica — a condition marked by eating non-food items, heightening poisoning risks—in those with cognitive impairments that can lead to disruptive and impulsive behavior known as Picacisim.  Among the various NDD syndromes that are significantly affected by diet or picacism, there is a particular syndrome where this aspect is most pronounced and crucial: the Prader-Willi Syndrom (PWS).
Prader-Willi Syndrome is a genetic disorder identified by anomalies in the hypothalamic-pituitary axis, leading to initial symptoms of hypotonia and poor weight gain, followed by hyperphagia and potential pathological obesity. This condition affects around 1 in 25,000 newborns. The disorder was initially thought to undergo two main nutritional phases \cite{cassidy2009prader}. However, subsequent research by Cassidy and Miller indicates a more intricate progression, involving seven phases and sub-phases \cite{cassidy2012prader}. The critical phase of hyperphagia, beginning at approximately 4 years, is characterized by an increased focus on food and persistent food-seeking, driven by hypothalamic dysfunction. This can lead to dangerous behaviors, such as eating inedible objects and food theft, potentially resulting in severe obesity and other health complications. Therefore, controlling or regulating these eating behaviors is essential for patient well-being.

Moreover, NDD subjects may exhibit selective eating habits based on food characteristics, which can lead to malnutrition or obesity, sometimes necessitating intensive care such as tube feeding. Children with certain NDDs often exhibit also restricted eating patterns and preferences for high-calorie, low-nutrient foods, contributing to obesity risk \cite{mari2014food}. Literature confirms that especially children with Down syndrome are very often morbidly or severely obese \cite{o2018prevalence}. Medications used to treat NDD symptoms can increase appetite and decrease metabolism, further exacerbating the risk of obesity. This prominence is also rooted in the multifaceted impact that nutrition has on cognitive, emotional, and physical development. For instance, certain NDDs are often accompanied by sensory sensitivities and gastrointestinal issues, making dietary choices and habits particularly challenging \cite{sharp2013feeding}. 

Optimal food management in these contexts can mitigate symptoms, improve gut health, and enhance the effectiveness of other therapeutic interventions \cite{hyman2012nutrient}. Additionally, specific nutrients or dietary patterns may influence neurological development and function, suggesting that tailored nutritional strategies can be integral to managing NDDs \cite{rucklidge2016nutrition}. 

However, implementing these strategies requires comprehensive food education, tailored not only to the needs of the individuals with NDD but also to their caregivers and educators\cite{curtin2010prevalence}. This education encompasses understanding the nutritional requirements, developing skills to handle dietary restrictions or sensory aversions, and recognizing the links between diet, behavior, and symptom management \cite{johnson2008eating}. Therefore, in the landscape of NDDs, an informed and personalized approach to food management, underpinned by robust food education, is essential in fostering optimal development and enhancing the overall well-being of those affected.

In the sphere of education, technology plays a pivotal role, offering unique opportunities for what is termed as "stealth learning." This approach, as detailed in Sharp's research \cite{Sharp}, leverages technology to enhance emotional engagement and expand learning opportunities, making it particularly effective for those with mental disabilities to develop a range of skills. Gaming, due to its inherent appeal, is frequently used in this context. Recent experiences, such as those documented by Mazzone \cite{Mazzone}, affirm the success of using games in educational settings, especially for individuals with neurodevelopmental disorders who are often attracted to technological devices.

Virtual reality (VR) emerges as a promising tool in this domain. Strickland's work \cite{strickland1996virtual} emphasizes VR's benefits, including controlled input stimuli and personalized learning environments. Recent interest in VR's application in NDD, as explored by researchers like Yufang \cite{Yufang} and Garzotto \cite{Garzotto}, underscores its adaptability to the specific learning needs of this demographic.

Our research aims to discern the most effective interface, VR or tablet, for imparting food education to individuals with NDD. We developed GEA, a web application with dual-interface functionality suitable for immersive experiences through a head-mounted display (HMD) or non-immersive experience on a tablet. The effectiveness of these interfaces is evaluated by comparing pre- and post-exposure questionnaire results.

\section{Related Work}
In the specialized area of food education, diverse interactive games for touchscreen devices have been introduced. A notable example is the Food Pyramid game by Colorado State University \cite{Serrano}, designed to educate children about the five primary food groups and apply this knowledge to meal and snack planning, thus boosting self-efficacy. The game's structure includes various challenges centered around the food pyramid, with findings indicating that challenge-based games are more impactful than storyline-based ones.

On the Healthy Eating website \footnote{https://www.healthyeating.org/Healthy-Kids/Kids-Games-Activities}, a collection of educational computer games related to food education is available. One of these games, titled "My Plate Match Game," is designed to instruct users in categorizing various types of food for a well-balanced diet, emphasizing the recommended daily quantities. The primary drawback of this game lies in its absence of proper feedback, as it continues without error tracking until all objectives are correctly achieved. Furthermore, the website also offers another game specifically centered around promoting a healthy breakfast, titled "Power Up Your Breakfast." This game shares similarities with the previously mentioned one, where users begin by responding to questions regarding the significance of a nutritious breakfast. Additionally, there is "My Very Own Pizza," a game that enables users to create a customized pizza without a specific end goal, allowing users to add any ingredients they desire.
While exploring the realm of food education games on the internet, we observed their prevalence, with another example being the Nourish Interactive website \footnote{http://www.nourishinteractive.com/kids/healthy-games}, which hosts a vast array of food-themed games.

Virtual reality technology has undergone significant development over the past half-century, starting from the creation of the Sensorama device in 1962, designed to provide an immersive film-watching experience, to the current devices that, although advanced, have not yet achieved complete sensory immersion \cite{biocca2013communication}.
Numerous applications of virtual reality technology have emerged over the years, prompting various experiments to assess their effectiveness and efficiency with persons with NDD. Some examples are the studies include those conducted by Lin (2016) \cite{lin2016augmented}, Lotan (2010) \cite{lotan2010virtual}, and Yalon-Chamovitz (2008) \cite{yalon2008virtual}. 

In Lin's study, 21 students with various disabilities engaged with an application comprising six games designed to enhance their understanding of geometry. The results were promising, as the repeated use of the application led to improved learning outcomes, reflected in increased success rates \cite{lin2016augmented}. In Lotan's research, a group of 44 adults, divided into two groups and affected by different intellectual and developmental disabilities (IDD), explored a virtual reality application aimed at supporting their physical fitness. However, the findings did not yield statistically significant results. \cite{lotan2010virtual}In the final example, 32 participants divided into two groups, all affected by various intellectual disabilities and severe cerebral palsy, interacted with a virtual reality application simulating leisure experiences and subsequently provided feedback through a questionnaire to assess their level of success and enjoyment \cite{yalon2008virtual}.

In the context of food education, VR applications have the potential to revolutionize how individuals acquire knowledge about nutrition, cooking techniques, and food-related concepts. Following, we provide a  selection of innovative VR-based food education games and experiences designed to engage learners while imparting essential culinary and nutritional insights.

One notable VR application is the "Cooking Simulator VR," which offers users a dynamic and interactive virtual kitchen environment. This simulation equips participants with the skills required to prepare a variety of dishes, including chopping ingredients, adhering to recipes, and monitoring the cooking process \cite{gorman2022using}. Real-time feedback enhances the learning process, making it an immersive platform for acquiring practical culinary expertise. The "Farm to Table VR" experience immerses users in the entire food production cycle. This includes farming and harvesting practices, animal husbandry, and an exploration of sustainable agricultural methods. Users gain insights into diverse cuisines and recipes while simultaneously learning about responsible and eco-friendly food production practices \cite{ruppert2011new}.

In the realm of nutrition education, the "Virtual Nutrition Adventure" VR application takes participants on a journey through the intricacies of the human digestive system. Users receive instruction on the significance of essential nutrients the process of food digestion, and are presented with opportunities to make informed dietary choices. Promoting food safety awareness, the "Food Safety VR" application immerses users in a virtual kitchen environment. Participants engage in interactive exercises to identify potential food safety hazards and acquire knowledge about proper food handling, storage, and hygiene practices \cite{davis2015virtual}.

In the "Virtual Food Science Lab," participants engage in experiments to understand food chemistry and explore the fundamental principles behind various cooking techniques, bridging the gap between scientific knowledge and culinary practice. The "Culinary Challenge VR" game challenges users to test and enhance their culinary skills in a competitive setting. Participants are required to prepare dishes under time constraints and follow specific instructions, fostering skill development, multitasking abilities, and culinary creativity \cite{karkar2018virtual}.

"Food World VR" offers an immersive journey around the world, enabling users to explore diverse cuisines, cultures, and food traditions. Through virtual visits to restaurants, markets, and street food stalls, participants gain insights into global food diversity, ingredient profiles, and cooking styles. The "Healthy Eating Quest VR" game encourages users to adopt healthy dietary choices through a virtual world filled with food-related challenges and puzzles. By navigating this immersive environment, participants gain an understanding of the importance of a balanced diet and the impact of food choices on overall health \cite{gorman2022using}.

The integration of immersive VR technology into food education has paved the way for engaging and informative learning experiences. However, based on our knowledge, very little has been done to develop a specific app for food education for individuals with NDD.

\section{GEA}
GEA ( acronym for "Gioco di Educazione Alimentare" - Food Education Game) is an immersive virtual reality mobile application running on smartphones and tablets and usable in virtual reality on Google Cardboard that aims to teach food education to individuals with NDD. GEA was the result of a codesign process carried out with therapists and specialists. The idea of food education activity arose because several studies have indicated that NDD patients frequently experience feeding difficulties \cite{MATSON2001165}. The activities contained in GEA, therefore seek to teach people with NDD very simple concepts concerning nutrition, such as i) at what time of day a certain food can be eaten, ii) which foods are healthier and which are not, iii) the allergens contained in a food.
\subsection{GEA Activities}
\subsubsection{Learn with the pyramid}
The objective of this game is to educate how to complete the food pyramid by selecting the appropriate foods for each level. A pyramid with five tiers displays in the virtual environment, with a pointer showing which level the user is currently finishing and a table containing three food possibilities. The pyramid consists of five levels, and the importance of foods for each level is based on the Mediterranean diet, so we used well-known foods (not only Italian but also belonging to other countries). At the lowest level (the largest), there are foods that have to be consumed daily, such as fruits, vegetables, and water. On the second level, there are carbohydrates and cereals, on the third level, there are meat and fish, on the penultimate level, dairy products such as cheese and yogurt, and on the final level (the smallest), foods that have to be consumed as little as possible, such as sweets and alcohol.
In order to provide a response, the user must select one of the foods on the table. If the answer is incorrect, a red circle appears, and the active level of the pyramid turns red; if the answer is accurate, the mascot appears with a joyful expression, and the level turns green. 

\subsubsection{Healthy or not?}
This game is intended to teach users to determine if a food is healthy or not. When the game begins, two dishes appear on the table of the virtual room, along with a bin and a visual explanation of how the game is played; the user must select the "junk food" and move it by maintaining their gaze on it until it is thrown into the bin. There are three repeats of this game, each with a new set of options, and after each response, similar feedback is displayed. This initially appeared to be a bad idea due to the thought of "throwing" away food, but it has been adopted and interpreted as a representation of the close correlation between junk foods and the trash can, so that the user may understand that these meals are highly unhealthy.

\subsubsection{Let's eat!}
This game aims to educate players on which is the best meal to eat a particular food. It was designed specifically for people who have trouble recognizing when they may and cannot consume certain foods. In this task, the virtual environment displays four images representing the four primary meals (breakfast, lunch, afternoon snack, and supper) as well as a dish; the objective is to select the correct meal/s during which the dish can be consumed. The selection can be done by staring at one of the meal options present on the wall.  Similar to the second game, there are three repeats, and feedback is provided after each answer. This game is based on the Mediterranean diet, thus there are specified items for each meal.

\subsubsection{Find the Allergens}
This activity aims to teach users about food allergens. The game environment is the same as the others; on the wall, there is a predefined number of allergens icons (4 for the first level of difficulty, 6 for the second, and 10 for the third) and food on the table.
The allergens included in the task are celery, clams, eggs, fish, dairy products, peanuts, shellfish, soy, nuts, and gluten. The task difficulty is based on the allergen itself and not on the type of food since some allergens are very common, like gluten, whereas others are very hard to detect, like celery. For this reason, in the Easy level, there are foods containing egg, fish, dairy products, and gluten, in the Medium one also soy and nuts. Finally, in the Difficult level also the remaining allergens.\\
The user must select all the allergens contained in the food. A countdown indicating the number of remaining allergens is also shown on the wall. Once the user has selected the correct number of allergens contained in the food, a confirm button appears. Then, the user receives the feedback back, showing also correct answers and wrong ones. 

\subsection{VR vs Touchscreen Version}
\begin{figure}[h!]
    \centering
    \includegraphics[width=1\textwidth]{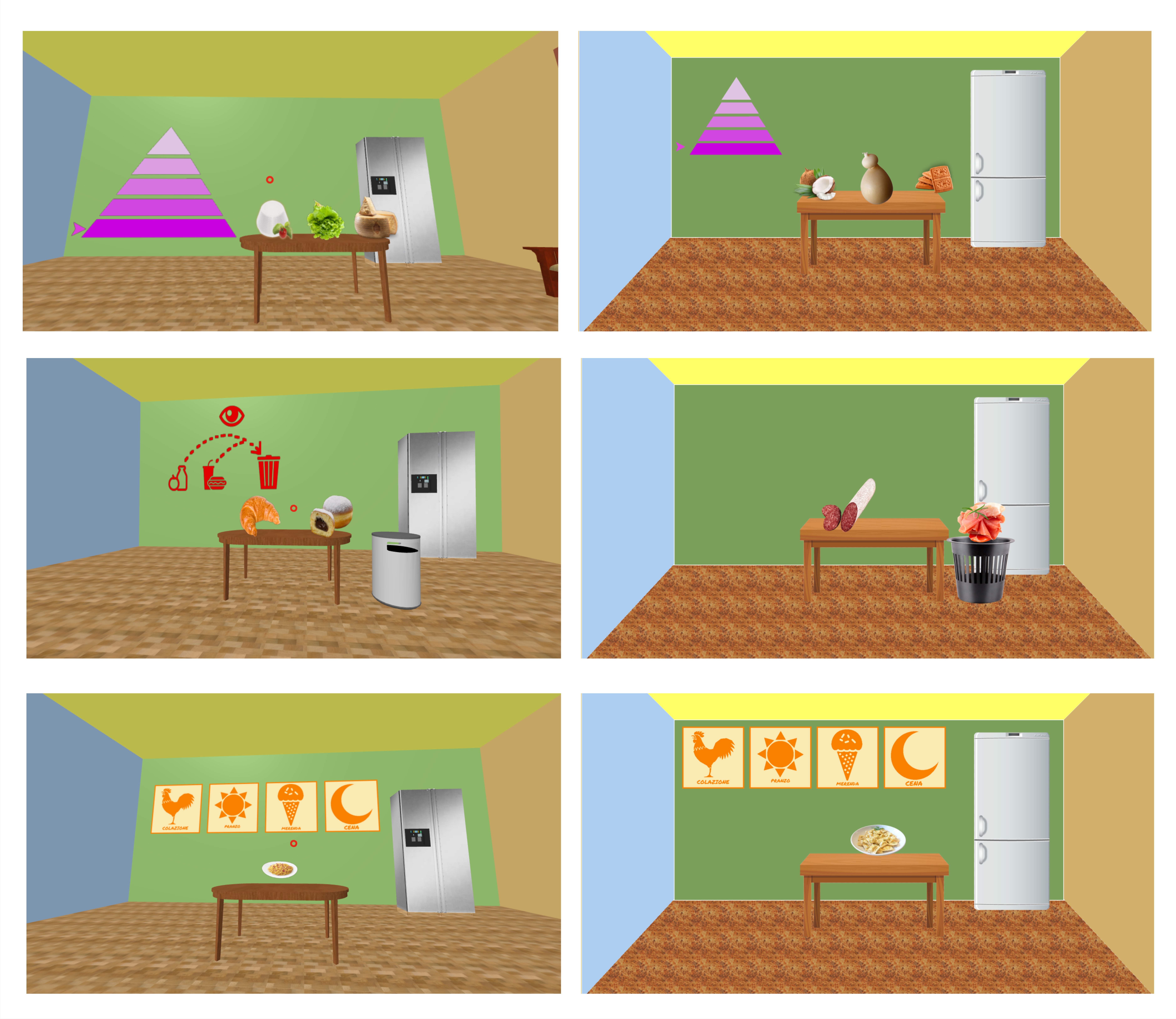}
    \caption{Comparison of each activity between the two versions of GEA}
    \label{fig:geapyramid}
\end{figure}

The two major differences there are between the versions of \textit{GEA} are related to user interaction and the design of the environment.
Concerning user interaction in the case of VR, it is based on gaze orientation and focus. By changing gaze direction, users update their view of the virtual environment and have the illusion of moving in different directions. Since, smartphone sensors (accelerometer and gyroscope) track only movement and orientation of the user’s head, the center of the smartphone screen is taken as a reference for the user's head. In case of the Tablet or Touchscreen interface, the user interaction is based on the Tap Gestures. 
The view of the environment is fixed, and the user can only interact with the food in the scene in order to give their answer. Figure\ref{fig:geapyramid} shows the difference between the two versions

\section{Study}
The study aimed to assess the potential of GEA through empirical research. Specifically, the objective was to compare the effectiveness of two different interfaces: Tablet (utilizing a touchscreen) and Virtual Reality. The study collected data on two key variables: 
\begin{itemize}
    \item The time interval it took patients to complete each mini-game at different levels of difficulty.
    \item The number of errors made during the gameplay.
\end{itemize}
To evaluate the effectiveness of these interfaces, a pre-test and a post-test were conducted to measure the learning delta related to food education.

\subsection{Participants}
An initial group of 45 participants with NDD was initially enrolled for the pre-test. The purpose of the pre-test was to determine which participants to include based on their existing knowledge of food education. The pre-test had a maximum score of 21 points. Participants with scores exceeding 17 were excluded from the study. Consequently, 20 participants remained eligible and took part in the research. Subsequently, these participants were randomly allocated to one of two conditions, namely Condition 1 (VR) or Condition 2 (Tablet). Participant information is detailed in the table below.\ref{table:selection}.

\begin{table}[h!]]
    \centering
    \begin{tabular} {|c|c|c|c|c|}
    \hline
       \textbf{ID} & \textbf{Age} & \textbf{Gender} & \textbf{Diagnosis} & \textbf{Condition}  \\\hline
       FEA49 & 18 & M & Prader-Willi & 2 \\\hline
       FEA18 & 26 & M & Cognitive delay and epilepsy & 2 \\\hline
       FEA55 & 32 & M & Down syndrome & 2\\\hline
       FEA41 & 26 & M & Autistic & 2\\\hline
       FEA30 & 18 & M & Cognitive delay, epilepsy and  encephalopathy & 2\\\hline
       FEA23 & 29 & F & Autism and cognitive delay & 2\\\hline
       FEA51 & 32 & F & Cognitive delay and epilepsy & 2\\\hline
       FEA19 & 29 & F & Medium-high cognitive delay & 2\\\hline
       FEA62 & 33 & M & High mental insufficiency & 1\\\hline
       FEA63 & 19 & M & Medium-high cognitive delay & 1\\\hline
       FEA52 & 27 & M & Cognitive delay & 1\\\hline
       FEA17 & 26 & M & Autism & 1\\\hline
       FEA44 & 27 & F & Low cognitive delay & 1\\\hline
       FEA11 & 37 & F & High mental insufficiency, psychosis & 1\\\hline
       FEA10 & 20 & F & Down syndrome & 1\\\hline
       FEA20 & 34 & F & Encephalopathy & 1\\\hline
       P4 & 51 & M & High cognitive delay & 2\\\hline
       P3 & 60 & M & Medium cognitive delay & 1\\\hline
       P6 & 59 & M & Medium cognitive & 1\\\hline
       P5 & 68 & F & High cognitive delay & 2\\\hline
     \end{tabular}
    \caption{Selected patients information \label{table:selection}}
    \end{table}
    
\subsection{Procedure}
The evaluation took place at two therapeutic centers in Milan, and the experimental study spanned three weeks. During each week of the study, participants had the opportunity to engage with all three tasks of GEA. The difficulty level of the games was systematically increased throughout the study. Before the commencement of the study, all participants responded to a 21-item questionnaire designed to assess their awareness regarding food and food education. The questionnaire was divided into three sections, each comprising seven questions, to cover the content of the three games offered by the application evenly. The questionnaire items are in the table below (Table\ref{tab: questionnaire}). In the first week of the study, all games were set to the "Easy" difficulty level. The games were adjusted to a "Medium" difficulty level during the second week. In the third week, the games were more challenging at the "Difficult" level.
An additional fourth week was introduced to accommodate users who were absent during one of the previous sessions or had started the evaluation one week later than the initial group. The specific version of the tasks varied depending on the condition assigned to each participant.
Following the three weeks of training, a post-test was conducted to assess the participants' knowledge of food education after their engagement with the GEA application using the same questionnaire with slightly different questions to avoid a learning effect. The post-test used the same test format as the pre-test and was administered to all 20 participants.
The GEA sessions occurred in a designated room at the daycare centres. We specifically selected quiet rooms to minimize ambient noise during recording. The furniture was minimal to reduce distractions. During the session, subjects sat in front of the researcher and wore the VR headset or played with the tablet. Participants could interrupt the experience whenever they wanted and/or request a break from the experimenter. The sessions lasted approximately 20 minutes.

\begin{table}[]
\resizebox{\textwidth}{!}{%
\begin{tabular}{|l|l|}
\hline
\textbf{Task} & \textbf{Question} \\ \hline
\multirow{7}{*}{Learn with Pyramid} & Do you know what is the food pyramid? a)Yes b)No \\ \cline{2-2} 
 & On the pyramid you put: a)Foods b)Seasons \\ \cline{2-2} 
 & Pyramid’s base represents foods that must be eaten.. a)Everyday b)Sometimes\\ \cline{2-2} 
 & What foods should be on the top of the pyramid? a)Sweet b)Vegetables \\ \cline{2-2} 
 & Pyramid’s top represents foods that must be eaten... a)More b)Less \\ \cline{2-2} 
 & What is better to eat more often? a)Meat b)Fried foods\\ \cline{2-2} 
 & How many steps does the pyramid have? a)5 b)7\\ \hline
\multirow{7}{*}{Healty or Not} & What is more healthy between fried fish and boiled one? a)Fried b)Boiled\\ \cline{2-2} 
 & Is it a good habit to eat every lunch at the fast-food? a)Yes b)No\\ \cline{2-2} 
 & Select the healthy food among the ones below: a)Apple b)Sweets\\ \cline{2-2} 
 & It is good to eat some fruit every day. a)True b)False\\ \cline{2-2} 
 & Fried chips are healthy. a)True b)False\\ \cline{2-2} 
 & What is less healthy between plain fish and fried one? a)Plain b)Fried\\ \cline{2-2} 
 & What is better between water and chocolate? a)Water b)Chocolate\\ \hline
\multirow{7}{*}{Let's Eat} & Can you eat pizza at breakfast? a)Yes b)No\\ \cline{2-2} 
 & Vegetables soup is often eaten at... a)Breakfast b)Dinner\\ \cline{2-2} 
 & Can foods be divided into groups based on daytime? a)Yes b)No \\ \cline{2-2} 
 & At lunch you eat: a)Pasta b)Milk with grains\\ \cline{2-2} 
 & At dinner you would eat: a) Meat and vegetables b)Snack and fruit juice\\ \cline{2-2} 
 & Is milk good at breakfast? a)Yes b)No\\ \cline{2-2} 
 & Is meat good for a snack? a)Yes b)No\\ \hline
\end{tabular}%
}
\caption{Questions in Pre-test and Post-test}
\label{tab: questionnaire}
\end{table}
\section{Results and Discussion}
Table \ref{table:posttest} presents the pre-test and post-test outcomes for each participant. An additional column has been included to compute the delta score, representing the difference between the scores from the two sessions. It is noteworthy that, in every instance, there was an improvement or no decrease in scores between the pre-test and post-test. The delta score ranged from a minimum of 2 to a maximum of 9.
  \begin{table}[ht]]
    \centering
    \begin{tabular} {|c|c|c|c|}
    \hline
       \textbf{ID} & \textbf{Pre-test Results} & \textbf{Post-test Results} & \textbf{Delta Score}\\\hline
       FEA49 & 17/21\footnotemark[1] & 19/21 & 2\\\hline
       FEA18 & 15/21 & 19/21 & 4\\\hline
       FEA55 & 11/21 & 16/21 & 5\\\hline
       FEA41 & 14/21 & 19/21 & 5\\\hline
       FEA30 & 16/21 & 20/21 & 4\\\hline
       FEA23 & 11/21 & 17/21 & 6\\\hline
       FEA51 & 8/21 & 17/21 & 9\\\hline
       FEA19 & 13/21 & 19/21 & 6\\\hline
       FEA62 & 15/21 & 18/21 & 3\\\hline
       FEA63 & 13/21 & 18/21 & 5\\\hline
       FEA52 & 13/21 & 20/21 & 7\\\hline
       FEA17 & 12/21 & 20/21 & 8\\\hline
       FEA44 & 16/21 & 20/21 & 4\\\hline
       FEA11 & 15/21 & 21/21 & 6\\\hline
       FEA10 & 16/21 & 21/21 & 5\\\hline
       FEA20 & 10/21 & 19/21 & 9\\\hline
       P4 & 15/21 & 20/21 & 5\\\hline
       P3 & 15/21 & 19/21 & 4\\\hline
       P6 & 15/21 & 20/21 & 5\\\hline
       P5 & 16/21 & 20/21 & 4\\\hline
     \end{tabular}
    \caption{Pre-test and Post-test results \label{table:posttest}}
    \end{table}

The data collected underwent analysis using SPSS (Statistical Package for Social Science). In the initial step, an assessment was conducted to ascertain the normality of the distribution. This involved the computation of values for skewness and kurtosis. The skewness index serves to determine the symmetry of data distribution in relation to a particular value, characterized as follows: a positive value indicates right skewness, where data points tend to cluster toward the lower end with an extended tail toward higher values, causing the distribution graph to stretch rightward. A negative value indicates left skewness, where the opposite situation prevails, causing the distribution graph to stretch leftward.
The kurtosis index, on the other hand, indicates the degree of deviation from a normal distribution, thereby signifying the relative weight of values located at the distribution's tails compared to those in its central region.
Table \ref{table:aak} provides evidence that the distribution adheres to normality based on the aforementioned values. This confirms the suitability for conducting inferential analysis despite the relatively small sample size. Regarding the results obtained in the pre-test and post-test, the average score increased from 13.80 to 19.10, the minimum score doubled (from 8 to 16), and the maximum attainable score (21) was also reached in the post-test.

\begin{table}[ht]
    \centering
    \begin{tabular} {|c|c|c|}
    \hline
       & Pre-test & Post-test \\\hline
       Population & 20 & 20  \\\hline
       Average & 13,80 & 19,10\\\hline
       Standard deviation & 2,375 & 1,334\\\hline
       Asymmetry & -0,940 & -0,793 \\\hline
       Asymmetry's standard error & 0,512 & 0,512 \\\hline
       Kurtosis & 0,286 & 0,223\\\hline
       Kurtosis' standard error & 0,992 & 0,992\\\hline
       Minimum & 8 & 16\\\hline
       Maximum & 17 & 21\\\hline
    \end{tabular}
    \caption{Average, asymmetry and kurtosis \label{table:aak}}
\end{table}
In general, all participants exhibited a notable increase in their post- questionnaire scores, demonstrating an average improvement of 46\%. This improvement was consistently observed in both groups, with the VR group showing a 42\% increase and the Tablet group showing a 41\% increase. Specifically, the results within the VR condition indicated a statistically significant difference (t=-9.333; df=9; p<.01) between the pre-test score (M=14) and the post-test score (M=19.6). Similarly, in the Tablet condition, there was a statistically significant difference (t=-8.660; df=9; p<.01) between the pre-test score (M=13.6) and the post-test score (M=18.6). An unpaired t-test was conducted to explore whether there was a significant difference between the VR and Tablet conditions in improving participants' performance. The results showed no significant difference (t=0.81; df=18; p=0.93) between the two conditions. Our study did not yield evidence to support the superiority of one interface over the other in enhancing knowledge. These results affirm the effectiveness of GEA as a valuable learning tool for individuals with NDD. Both the VR and Tablet interfaces demonstrated suitability for our target population.
\\
\section{Conclusions}
The target user group of GEA primarily consists of individuals with neurodevelopmental disorders. These individuals often require more extensive support, even in nutrition education, given their daily cognitive challenges. 
GEA proved to be an effective learning tool, with tested patients exhibiting an increase in their scores on general knowledge about food after engaging with the game, demonstrating a positive outcome. Generally, there was no significant difference in terms of improvement between the two versions of the game. However, our data indicated that virtual reality performed particularly well with individuals experiencing high cognitive delay and autism disorders, suggesting its potential benefits in these specific cases.
The results of our study do not provide significant evidence in favor of a VR interface over a tablet interface. This finding is particularly interesting in a context where there has been a rush towards new and highly advanced technologies as the sole means of improving learning capabilities, especially in recent times and in the field of education. However, our study demonstrates that this assertion is not necessarily true. In our study, the improvement achieved was comparable between the two interfaces, without the more advanced VR interface proving superior to the more traditional and commonly adopted tablet interface. It's crucial to recognize that not all advanced technologies necessarily represent an advancement in every context. In education, particularly, choosing the most suitable tool should take precedence over pursuing the latest and most cutting-edge solution. The effectiveness of an educational tool depends on its alignment with the learning objectives and the needs of the students. Therefore, educators should carefully evaluate whether a technological innovation enhances the learning experience and outcomes before adopting it. This is particularly relevant in the NDD arena.

\subsection{Limitation and Future Work}
In our evaluation, we encountered certain limitations, including the relatively small number of participants. A larger sample size could have generated more robust results, potentially revealing differences in performance between the two technology interfaces. Another constraint was the limited time available for experimentation. We could only assess GEA over a three-week period, and a more extended period of use might have led to even greater learning outcomes.
As part of our future work, we have several ideas to expand and enhance the application. First, conduct a more extensive and long-term evaluation of the application to gain a deeper understanding of its potential for improving learning outcomes including an evaluation of the fourth game, "Find the allergen!" which has not been tested yet. Second, develop additional games related to food and nutrition, such as helping users understand and balance their intake of carbohydrates, sugars, proteins, and fats, or suggesting suitable seasonal food choices.

%
%
%
\bibliographystyle{splncs04}
\bibliography{main}

\end{document}